\documentclass[10pt, aps, prd, amsmath, amssymb, floats, floatfix, twocolumn, notitlepage, superscriptaddress, nofootinbib, showpacs, longbibliography]{revtex4-2}
\usepackage[T1]{fontenc}
\usepackage[utf8]{inputenc}
\usepackage{lmodern}
\usepackage{xcolor}
\usepackage{mathbbol}
\usepackage{verbatim}
\usepackage{physics}
\usepackage{orcidlink}
\definecolor{linkcolor}{rgb}{0.6, 0.0, 0.0}
\usepackage{graphicx}
\usepackage[all]{hypcap}
\usepackage{xspace}
\usepackage{amssymb}
\usepackage{amsmath}
\usepackage[normalem]{ulem}
\usepackage{bm}
\usepackage{siunitx}
\usepackage{microtype}
\usepackage[english]{babel}
\usepackage{blindtext}
\usepackage{array}
\usepackage{tabularx}
\usepackage{multirow}
\usepackage{subfigure}
\usepackage{natbib}
\graphicspath{{figs/}}

\usepackage{cleveref}
\crefname{section}{Section}{Sections}
\crefname{equation}{Eq.}{Eqs.}
\crefname{figure}{Fig.}{Figs.}
\crefname{appendix}{Appendix}{Appendices}
\usepackage{soul}
\allowdisplaybreaks

\definecolor{db}{rgb}{0.0, 0.0, 0.62}
\definecolor{dm}{rgb}{0.7, 0.01, 0.7}
\definecolor{dr}{rgb}{0.55, 0.0, 0.0}

\newcommand{\Mp}{M_\mathrm{P}}

\newcommand{\SW}{\Delta_{\rm SWGC}}
\newcommand{\SSW}{\Delta_{\rm SSWGC}}

\begin{document}

\title{Swampland diagnostics and observational tests of the inflationary universe\\ in the Palatini formalism}

\author{Saeed Noori Gashti\orcidlink{0000-0001-7844-2640}}
\email{sn.gashti@du.ac.ir; saeed.noorigashti70@gmail.com}
\affiliation{School of Physics, Damghan University, P. O. Box 3671641167, Damghan, Iran}
\affiliation{Center for Theoretical Physics, Khazar University, 41 Mehseti Street, Baku, AZ1096, Azerbaijan}

\author{Mohammad Reza Alipour\orcidlink{0000-0001-8074-7865}}
\email{mr.alipour@stu.umz.ac.ir}
\affiliation{School of Physics, Damghan University, P. O. Box 3671641167, Damghan, Iran}
\affiliation{Center for Theoretical Physics, Khazar University, 41 Mehseti Street, Baku, AZ1096, Azerbaijan}
\vspace{1cm}
\author{Mohammad Ali S. Afshar\orcidlink{0009-0001-3133-5992}}
\email{m.a.s.afshar@gmail.com}
\affiliation{Department of Physics, Faculty of Basic Sciences, University of Mazandaran P. O. Box 47416-95447, Babolsar, Iran}
\affiliation{School of Physics, Damghan University, P. O. Box 3671641167, Damghan, Iran}
\affiliation{Center for Theoretical Physics, Khazar University, 41 Mehseti Street, Baku, AZ1096, Azerbaijan}

\author{Kazuharu Bamba\orcidlink{0000-0001-9720-8817}}
\email{bamba@sss.fukushima-u.ac.jp}
\affiliation{Faculty of Symbiotic Systems Science, Fukushima University, Fukushima 960-1296, Japan}

\begin{abstract}
We consider quintessential inflation in a quadratic Palatini model with a generalized Peebles-Vilenkin potential, where the same scalar field describes both the inflationary plateau and the later inverse-power-law regime. Passing to the Einstein frame gives a canonically normalized field and an explicit effective potential. The inflationary predictions are then reassessed, together with the further-refined de Sitter, scalar weak-gravity, strong scalar weak-gravity, distance, and trans-Planckian censorship conjectures. The value of the tensor-to-scalar ratio stays below the observational bounds used in this work, whereas the spectral index of scalar fluctuations changes with the potential exponent and with the duration of inflation. The further-refined de Sitter bound is satisfied only over part of the trajectory. At the pivot, both weak-gravity diagnostics are negative in the reduced-Planck-mass convention, although they become positive at finite field values. The field excursion is super-Planckian, but this does not by itself imply a failure of the distance conjecture; the trans-Planckian censorship criterion is not met in the assumed slow-roll regime.
\end{abstract}

\maketitle

\section{Introduction}
Inflationary model building is becoming more sensitive to relatively small shifts in the preferred value of the scalar spectral index. This is particularly important for models whose predictions already lie close to the region allowed by observations. For the combination of ACT DR6 with Planck, CMB-lensing, and BAO data, the preferred scalar tilt is on the higher side. The P-ACT-LB dataset gives \(n_\mathrm{s}=0.9743\pm0.0034\) at \(68\%\) confidence, while adding the BICEP/Keck 2018 data gives the representative bound \(r<0.038\) at \(95\%\) confidence. A later combination that also includes DESI DR2, commonly denoted by P-ACT-LB2, leads to \(n_\mathrm{s}\simeq0.9752\pm0.0030\) \cite{ACTDR6}. Thus, the preferred value of \(n_\mathrm{s}\) is not completely independent of the particular dataset combination.

In the present work, we use the P-ACT-LB result as a representative observational benchmark. We do not regard it as a unique constraint, and we are not attempting a full likelihood analysis. Its role here is simply to provide a fixed observational reference against which the predictions of the effective model can be compared \cite{ACTDR6,Planck2018,BK18}.

These numbers by themselves do not determine which inflationary model is realized in nature. They are nevertheless useful when testing modified-gravity scenarios. In pure Palatini \(F(R)\) gravity, the scalar field that appears in the scalar–tensor representation is auxiliary rather than dynamical. This is different from metric \(F(R)\) gravity, where the corresponding scalaron propagates as an independent degree of freedom. The situation can change when additional couplings to matter or derivatives are introduced, since these terms may alter the dynamics of the scalar sector. In the quadratic \(F(R,X)\) model studied in this work, the Einstein-frame action contains a canonical kinetic term for the scalar field, while the nonlinear curvature contributions are incorporated into its effective potential. The \(F(R,X)\) theory considered here has a somewhat different organization. The curvature dependence and the scalar kinetic invariant enter together already in the original action, and, for the quadratic realization studied here, the Einstein-frame description reduces to ordinary Einstein gravity plus a canonically normalized scalar field. The nonlinear contributions inherited from the modified curvature sector are then encoded in the resulting effective potential rather than in a non-standard gravitational kinetic term.

This effective potential is well suited to a quintessential-inflation setup because its two asymptotic regions have rather different roles. For large negative field values, the potential tends toward an approximately constant plateau. On the positive-field side, it can retain a decreasing inverse-power tail. The same canonical scalar can therefore account for the inflationary phase and, at much later times, enter a quintessence-like regime within one Einstein-frame effective description. This gives a simple way to track the scalar evolution from inflation through the later dark-energy phase without introducing a separate scalar sector \cite{Dimopoulos2025,Dioguardi2024}.

We use the Palatini formulation to investigate whether its different gravitational dynamics can make this particular quintessential-inflation model more consistent with observations. In the model considered here, the quadratic curvature term modifies the effective potential and creates a flatter region suitable for inflation, while preserving the inverse-power form required for the subsequent quintessence phase. This change in the potential can affect both the scalar spectral index and the tensor-to-scalar ratio, and thus alter the predictions for the cosmic microwave background. We examine these predictions in light of the scalar spectral index indicated by the P-ACT-LB data combination and the available bounds on tensor modes. The purpose is to assess the viability of this specific Palatini realization, not to argue that ACT DR6 requires the Palatini formulation. A claim that it is generally preferable to the metric approach would need a separate, quantitative comparison between the two formulations.

There is also a broader motivation for testing this construction against the Swampland program. The underlying idea is that a low-energy effective field theory may not be compatible with every possible ultraviolet completion involving quantum gravity. In this terminology, effective theories that can arise from a consistent quantum-gravitational completion are associated with the landscape, whereas those that cannot are placed in the swampland \cite{k,o,bbb}. A number of conjectures have been proposed in an attempt to turn this general idea into quantitative criteria. In cosmology, the Weak Gravity Conjecture (WGC), the de Sitter (dS) conjecture and its further-refined form, together with the Trans-Planckian Censorship Conjecture (TCC), have received particular attention \cite{50,51,52,54,55,56,57,59,60,Palti2017,GonzaloIbanez2019,AndriotRoupec2019,Palti111,vanBeest:2022}. These conjectures are motivated by different aspects of effective field theory and quantum gravity. They should therefore be treated as distinct conditions rather than as different expressions of one universal statement. Considered together, they are better viewed as a set of complementary diagnostics for asking how a given low-energy cosmological model sits with respect to proposed quantum-gravitational consistency requirements.

Here we concentrate on the quadratic Palatini \(F(R,X)\) realization of quintessential inflation. The gravitational setup, its Einstein-frame reduction, and the corresponding inflationary benchmark were already derived in Ref.~\cite{Dimopoulos2025}. We therefore do not present that construction as a new result. Instead, we use the previously obtained effective theory as the starting point and ask how it behaves under a wider collection of Swampland criteria. We pay particular attention to the canonical field-space description, to scalar weak-gravity conditions formulated directly in terms of the potential, to the asymptotic limits of these quantities, and to the sensitivity of the inflationary predictions to the assumed number of e-folds. This separation is important when reading the results. Compatibility with CMB data tests the phenomenology of the model, whereas the Swampland conditions are conjectural statements about possible quantum-gravitational consistency. Passing one type of test does not imply passing the other. Recent observational developments have likewise motivated renewed analyses of inflationary models in a range of alternative theoretical frameworks; some representative examples can be found in Refs.~\cite{18',19',20',21',22',23',24',25',26',27',28',29',30',31',32',33',34',35',37',37a}.

More generally, the theoretical and cosmological aspects of \(f(R)\) and extended-gravity theories, including their metric, Palatini, and metric-affine formulations, have been reviewed extensively in the literature \cite{Sotiriou:2008rp,DeFelice:2010aj,Nojiri:2010wj,Clifton:2011jh,Capozziello:2011et,Nojiri:2017ncd,Kallosh:2025ijd,Weinberg:2008,Weinberg:2009,Bamba:2015,Bamba:2016,Palti1111}. In particular, the role of an \(R^2\) term in Palatini inflation with additional scalar fields has been studied in specific models \cite{Dioguardi2024}. For a review of inflationary dynamics and reheating, see Ref.~\cite{Bassett2006}. The Starobinsky model also provides a foundational example of inflation driven by higher-curvature effects \cite{Starobinsky1980}.

The rest of the paper is arranged as follows. Section~\ref{sec:formalism} sets up the Palatini \(F(R,X)\) theory, introduces the auxiliary-field formulation, and derives the Einstein-frame action and its exact effective potential. In Sec.~\ref{sec:PV}, we specialize this construction to the generalized Peebles-Vilenkin potential and derive the exact field-derivative relations used later. Section~\ref{sec:inflation} is devoted to the inflationary regime. We discuss the slow-roll quantities, the relation between the field and the number of e-folds, the large-\(\alpha\) plateau regime, and the leading contribution to the tensor-to-scalar ratio. The further refined de Sitter conjecture, together with the scalar weak-gravity and strong scalar weak-gravity criteria, is written in canonical Einstein-frame variables in Sec.~\ref{sec:swampland}. Their asymptotic behaviour is then examined in Sec.~\ref{sec:limits}, both on the inflationary plateau and along the inverse-power quintessence tail. The distance conjecture and the trans-Planckian censorship conjecture are treated independently in Secs.~\ref{sec:distance} and \ref{sec:TCC}. In Sec.~\ref{sec:numerics}, we solve the model numerically and compare the resulting inflationary observables with the representative CMB constraints. The FRDSSC, SWGC, SSWGC, distance, and TCC diagnostics are also evaluated at the inflationary pivot. A combined comparison between these theoretical criteria and the observational results is collected in Sec.~\ref{sec:summarytable}. Section~\ref{sec:kination} follows the scalar evolution beyond inflation, from the inflationary solution through kination and onward to the quintessence tail. The physical interpretation of the results, as well as the main assumptions and limitations of the analysis, is discussed in Sec.~\ref{sec:discussion}. We then close with a summary of the main conclusions. The appendices contain the longer derivative formulae for the inflationary and quintessence branches, additional details of the slow-roll benchmark, the observational form of the FRDSSC conditions, the asymptotic no-go results, and the numerical treatment of the TCC.
\section{Palatini $F(R,X)$ gravity and the Einstein frame}
\label{sec:formalism}
The reason for using the quadratic Palatini form is rather simple. After the reduction, the scalar kinetic term has the usual canonical form, but the potential is no longer the original Jordan-frame one. For the Peebles-Vilenkin construction this modification is enough to flatten the negative-field branch. The positive-field branch, on the other hand, keeps its inverse-power falloff and can still be used for the quintessence stage \cite{Dimopoulos2025}.
We take the Jordan-frame action in the form \cite{Dimopoulos2025}
\begin{equation}
S_\mathrm{J}=\int d^4x\sqrt{-g_\mathrm{J}}\left[\frac{\Mp^2}{2}F(R_\mathrm{X})-V_\mathrm{J}(\varphi)\right],
\label{eq:actionJ}
\end{equation}
where
$
M_\mathrm{P}=(8\pi G)^{-1/2}
$
is the reduced Planck mass. The scalar field $\varphi$ and the scale $M$
have mass dimension one, while the quadratic Palatini parameter has
$\alpha=2\times10^8M_\mathrm{P}^{-2}$. Numerical calculations are performed in
reduced Planck units, $M_\mathrm{P}=1$.
We define
\begin{equation}
\begin{split}
&R_\mathrm{X}\equiv R+X,
\qquad
X\equiv-\frac{1}{M_\mathrm{P}^2}
g_\mathrm{J}^{\mu\nu}\partial_\mu\varphi\partial_\nu\varphi,\\
&R=g_\mathrm{J}^{\mu\nu}R_{\mu\nu}(\Gamma).
\end{split}
\label{eq:RX}
\end{equation}
With these definitions, $R_\mathrm{X}$ behaves as a curvature scalar and has the
same dimensional properties as the usual scalar curvature. In fact, after
performing the transformation from the Jordan frame to the Einstein frame,
the kinetic part of the scalar field obtains the canonical form, which will
be used in the rest of our calculations.
Now, in order to rewrite the curvature-dependent part of the action in a more
suitable way, we introduce an auxiliary field $\zeta$. By using this field,
the curvature terms can be linearized and Eq.~\eqref{eq:actionJ} takes the
following form:
\begin{equation}
\begin{split}
S_\mathrm{J}=&\int d^4x\sqrt{-g_\mathrm{J}}\bigg\{\frac{\Mp^2}{2}
\left[F(\zeta)+F_{,\zeta}(\zeta)(R_\mathrm{X}-\zeta)\right]\\
&\hspace{42pt}-V_\mathrm{J}(\varphi)\bigg\}.
\end{split}
\label{eq:auxiliary}
\end{equation}
Varying $\zeta$ gives back the original theory whenever $F_{,\zeta\zeta}\neq0$. We then introduce the Einstein-frame metric
\begin{equation}
g^\mathrm{E}_{\mu\nu}=\Omega^2g^\mathrm{J}_{\mu\nu},
\qquad
\Omega^2=F_{,\zeta}.
\label{eq:conf}
\end{equation}
The resulting action reads
\begin{equation}
\begin{split}
S_\mathrm{E}=&\int d^4x\sqrt{-g_\mathrm{E}}\bigg[
\frac{\Mp^2}{2}R_\mathrm{E}
\\
&-\frac12 g_\mathrm{E}^{\mu\nu}\partial_\mu\varphi\partial_\nu\varphi
-U(\zeta,\varphi)\bigg],
\end{split}
\label{eq:actionE}
\end{equation}
with
\begin{equation}
U(\zeta,\varphi)=
\frac{V_\mathrm{J}(\varphi)}{F_{,\zeta}^2}
-\frac{\Mp^2F(\zeta)}{2F_{,\zeta}^2}
+\frac{\Mp^2\zeta}{2F_{,\zeta}}.
\label{eq:Uaux}
\end{equation}
The remaining auxiliary constraint takes the form
\begin{equation}
\frac14\left(2F(\zeta)-\zeta F_{,\zeta}(\zeta)\right)
=\frac{V_\mathrm{J}(\varphi)}{\Mp^2}.
\label{eq:auxconstraint}
\end{equation}
This constraint is enough to eliminate $\zeta$ from the potential, giving
\begin{equation}
U(\zeta)=\frac{\Mp^2\zeta}{4F_{,\zeta}(\zeta)}.
\label{eq:Uzeta}
\end{equation}
For the calculation that follows, this is the required Einstein-frame form. It is the same reduction used in the quadratic Palatini quintessential-inflation construction of Ref.~\cite{Dimopoulos2025}.
For the quadratic choice of $F$, the auxiliary equation simplifies immediately:
\begin{equation}
F_{,\zeta}=1+2\alpha\zeta,
\qquad
\frac14\left(2F-\zeta F_{,\zeta}\right)=\frac{\zeta}{4},
\end{equation}
so that
\begin{equation}
\zeta(\varphi)=\frac{4V_\mathrm{J}(\varphi)}{\Mp^2}.
\label{eq:zetaV}
\end{equation}
Substituting this result back into the Einstein-frame potential gives the central relation of the model,
\begin{equation}
U(\varphi)=
\frac{V_\mathrm{J}(\varphi)}{1+8\alpha V_\mathrm{J}(\varphi)/\Mp^2},
\label{eq:Umaster}
\end{equation}
provided
$1+8\alpha V_\mathrm{J}/\Mp^2>0$.
We therefore do not introduce another scalar variable. The field $\varphi$ in Eq.~\eqref{eq:actionE} is already canonical, and the derivatives used later are simply taken with respect to $\varphi$.

\section{Generalized Peebles-Vilenkin potential}
\label{sec:PV}
We now choose the generalized Peebles-Vilenkin potential
\begin{equation}
V_\mathrm{J}(\varphi)=
\begin{cases}
\lambda^k(\varphi^k+M^k), & \varphi\leq0,\\[5pt]
\displaystyle
\lambda^k\frac{M^{k+q}}{\varphi^q+M^q}, & \varphi>0,
\end{cases}
\label{eq:PVpotential}
\end{equation}
Here $k$ takes even positive values in our analysis. The first branch of the
potential, which is responsible for inflation, is fixed by this parameter.
The parameter $M$ is introduced at the point where the two parts of the
potential are connected. On the other hand, the value of $q$ affects the
inverse-power term that appears for $\varphi>0$ and becomes important after
the transition region.
It is also necessary to check the behavior at the matching point. This is
because the SSWGC contains the fourth derivative of the potential, and this
derivative can be sensitive to the way that the two branches are connected.
For different choices of $k$ and $q$, the differentiability order of the
potential can change. In particular, when $k=q=4$,
\[
V''''(0^-)=+24\lambda^4,
\qquad
V''''(0^+)=-24\lambda^4,
\]
whereas all derivatives through third order agree at $\varphi=0$. The joined potential is consequently $C^3$, but it is not $C^4$. Since $V'=V''=V'''=0$ there, the same jump appears in $U''''$. More generally, when $k=q$ the joined potential is $C^{k-1}$ at the transition. Thus the SSWGC is evaluated branch by branch wherever the fourth derivative exists. For $k=q=2$ and $4$ the point $\varphi=0$ itself is omitted from the SSWGC because $U''''$ is not defined there; no finite interval is removed.
The Einstein-frame potential becomes
\begin{equation}
U_-(\varphi)=
\frac{\lambda^k(\varphi^k+M^k)}
{1+8\alpha\lambda^k(\varphi^k+M^k)/\Mp^2},
\quad \varphi\leq0,
\label{eq:Uminus}
\end{equation}
and
\begin{equation}
U_+(\varphi)=
\frac{\lambda^kM^{k+q}/(\varphi^q+M^q)}
{1+8\alpha\lambda^kM^{k+q}/[\Mp^2(\varphi^q+M^q)]},
\hspace{3pt} \varphi>0.
\label{eq:Uplus}
\end{equation}
The asymptotic limits are
\begin{equation}
\begin{split}
U_-(\varphi\rightarrow-\infty)&=\frac{\Mp^2}{8\alpha},\\
U_+(\varphi\rightarrow+\infty)&\sim
\lambda^kM^{k+q}\varphi^{-q}.
\end{split}
\label{eq:asymptoticU}
\end{equation}
The limiting forms already tell us how the two sides of the potential behave. For large negative $\varphi$ the potential tends to $M_\mathrm{P}^2/(8\alpha)$. For positive $\varphi$ it falls as an inverse power. The plateau is therefore set by $\alpha$, while the exponent $q$ determines the eventual tail.
\begin{figure}[t]
\centering
\includegraphics[width=\linewidth]{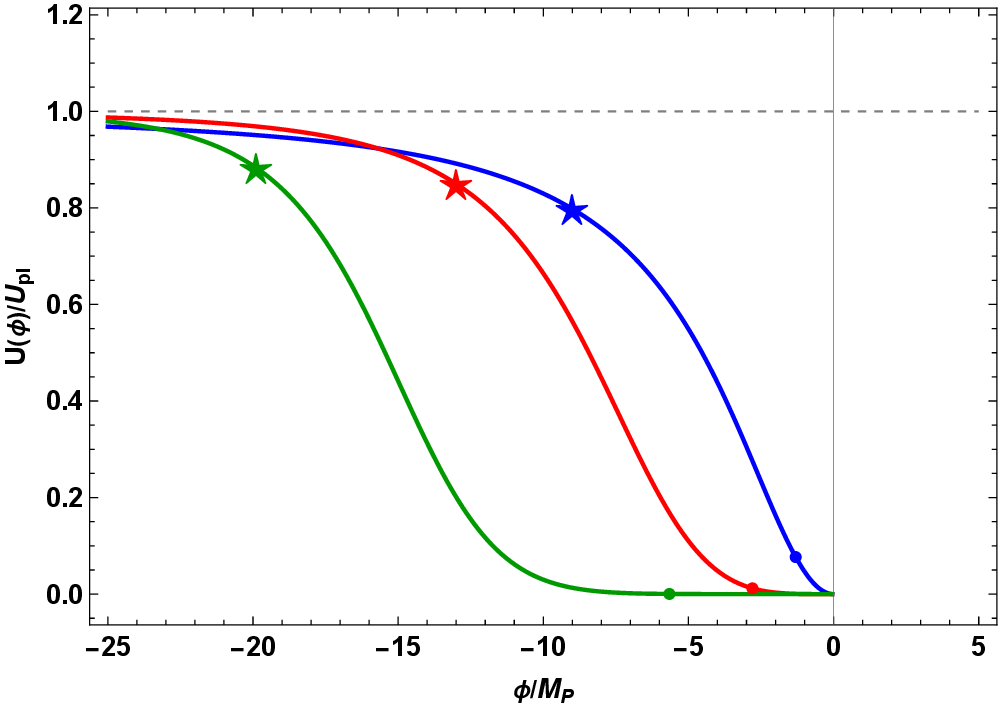}
\caption{We plot the scalar weak-gravity diagnostics for the canonical Einstein-frame potential at \(k=2,4,8\), taking \(q=k\) for each case. The plots include both branches of the potential. The entire relevant field domains are displayed. For $k$ equals $q$ equals $2$ and $4$ the only matching point $\varphi=0$ is omitted from the derivative test because $U''''$ is undefined there; no finite neighborhood is removed. The CMB pivot points are marked on the branch. Finite positive SWGC intervals occur during the evolution and then negative values appear at field distance.}
\label{fig:potential}
\end{figure}
In Fig.~\ref{fig:potential} the plateau height is the same for the three benchmark values of $k$, since it is set by $\alpha$. What changes with $k$ is the way the potential approaches that plateau. The numerical pivot and the end of inflation consequently move to more negative values as $k$ is increased.
\subsection{Exact derivatives}
Define
\begin{equation}
C\equiv\frac{8\alpha}{\Mp^2},
\qquad
D\equiv1+C V_\mathrm{J}.
\label{eq:CD}
\end{equation}
From $U=V_\mathrm{J}/D$ we obtain, without any further approximation,
\begin{align}
U^{\prime}&=\frac{V'}{D^2},
\label{eq:U1}\\
U^{\prime\prime}&=\frac{V''}{D^2}-\frac{2C(V')^2}{D^3},
\label{eq:U2}\\
U^{\prime\prime\prime}&=\frac{V'''}{D^2}-\frac{6CV'V''}{D^3}+
\frac{6C^2(V')^3}{D^4},
\label{eq:U3}\\
U^{\prime\prime\prime\prime}&=\frac{V''''}{D^2}-\frac{8CV'V'''}{D^3}
-\frac{6C(V'')^2}{D^3}
\nonumber\\
&\quad+\frac{36C^2(V')^2V''}{D^4}-\frac{24C^3(V')^4}{D^5}.
\label{eq:U4}
\end{align}
No expansion in $\alpha$ has been made in these derivatives. The expressions are the ones used directly in the finite-field SWGC and SSWGC calculations.
\section{Inflationary observables}
\label{sec:inflation}
We can now use the usual single-field slow-roll quantities:
\begin{equation}
\epsilon_\mathrm{V}=\frac{\Mp^2}{2}\left(\frac{U'}{U}\right)^2,
\qquad
\eta_\mathrm{V}=\Mp^2\frac{U''}{U}.
\label{eq:slowroll}
\end{equation}
At first order,
\begin{equation}
\begin{split}
&n_\mathrm{s}\simeq1-6\epsilon_V+2\eta_V,
\qquad
r\simeq16\epsilon_V,
\\&
A_\mathrm{s}\simeq\frac{U_*}{24\pi^2\Mp^4\epsilon_{V,*}}.
\label{eq:observables}
\end{split}
\end{equation}
The corresponding e-fold interval is
\begin{equation}
N_*\simeq\frac1{\Mp^2}
\int_{\varphi_{\rm end}}^{\varphi_*}
\frac{U}{U'}d\varphi,
\qquad
\epsilon_V(\varphi_{\rm end})=1.
\label{eq:Ngeneral}
\end{equation}
On the inflationary side we write $y=-\varphi>0$. With $M_\mathrm{P}=1$, the e-fold integral has a simple primitive for $k\neq2$:
\begin{equation}
\begin{split}
N_*=&\frac{1}{k}\Bigg[\frac{y^2}{2}
+\frac{M^k y^{2-k}}{2-k}\\
&+8\alpha\lambda^k\left(
\frac{y^{k+2}}{k+2}+M^k y^2
+\frac{M^{2k}y^{2-k}}{2-k}\right)\Bigg]_{y_{\rm end}}^{y_*}.
\end{split}
\label{eq:Nprimitive}
\end{equation}
For $k=2$ the same result is obtained by taking the logarithmic limit. The numerical points reported below are obtained from the full expression, not from this limiting approximation. The latter is used only to see the scaling and to compare with the earlier treatment \cite{Dimopoulos2025}.
For the numerical work it is also useful to write $\epsilon_\mathrm{V}$ and $\eta_\mathrm{V}$ directly in terms of the Jordan-frame potential:
\begin{equation}
\epsilon_\mathrm{V}=\frac{\Mp^2}{2}\frac{(V')^2}{V^2D^2},
\label{eq:epsJordan}
\end{equation}
and
\begin{equation}
\eta_\mathrm{V}=\Mp^2\left[\frac{V''}{VD}
-\frac{2C(V')^2}{VD^2}\right].
\label{eq:etaJordan}
\end{equation}
\subsection{Large-$\alpha$ plateau expansion}
In what follows, large $\alpha$ has a limited meaning. We require only
\begin{equation}
\frac{8\alpha V_\mathrm{J}}{M_\mathrm{P}^2}\gg1.
\end{equation}
This condition is used only for the expansion written below. It is not a claim about a UV cutoff, nor does it identify a strong-coupling scale.
In this regime the effective potential reads, to the displayed order,
\begin{equation}
U\simeq
\frac{M_\mathrm{P}^2}{8\alpha}
\left[1-\frac{M_\mathrm{P}^2}{8\alpha V_\mathrm{J}}
+\mathcal{O}\left(\frac{M_\mathrm{P}^4}{\alpha^2V_\mathrm{J}^2}\right)\right].
\label{eq:plateauexp}
\end{equation}
On the inflationary side, taking $y=-\varphi\gg M$ gives
\begin{equation}
U\simeq
\frac{M_\mathrm{P}^2}{8\alpha}
\left[1-\frac{M_\mathrm{P}^2}{8\alpha\lambda^k y^k}+\cdots\right].
\label{eq:plateauPower}
\end{equation}
The e-fold integral simplifies at the same order:
\begin{equation}
N_*\simeq
\frac{1}{kM_\mathrm{P}^2}
\left[\frac{y^2}{2}
+\frac{8\alpha\lambda^k}{M_\mathrm{P}^2}
\frac{y^{k+2}}{k+2}\right]_{y_{\rm end}}^{y_*}.
\label{eq:Nlargealpha}
\end{equation}
There is one more simplification if the second term in the bracket is the parametrically larger one:
\begin{equation}
N_*\simeq
\frac{8\alpha\lambda^k}
{k(k+2)M_\mathrm{P}^4}y_*^{k+2}.
\label{eq:Nstrong}
\end{equation}
Equation~\eqref{eq:Nstrong} is kept as an asymptotic guide. The benchmark numbers below come from the exact equations.
At fixed $A_\mathrm{s}$, the leading tensor estimate takes the simple form
\begin{equation}
r\simeq\frac{1}{12\pi^2\alpha M_\mathrm{P}^2 A_\mathrm{}s}.
\label{eq:rstrong}
\end{equation}
Putting the above values of $\alpha=2\times10^8M_\mathrm{P}^{-2}$ and
$A_\mathrm{s}=2.1\times10^{-9}$ in the approximate expression, we get
$r\simeq0.0201$. It is found that the value coming from the numerical
analysis is smaller than this estimate. The reason is that during the whole
inflationary evolution the field does not exactly follow the asymptotic
solution. In particular, this difference appears when the inflationary field
approaches the end of its evolution.
For the same limit, the scalar tilt is obtained as
\begin{equation}
n_\mathrm{s}\simeq1-\frac{2(k+1)}{(k+2)N_*},
\label{eq:nsstrong}
\end{equation}
while the finite-$\alpha$ terms are retained in the numerical solution.
\section{Swampland criteria in the canonical Einstein frame}
\label{sec:swampland}
We first fix the field-space convention used in the swampland tests. After the Einstein-frame reduction the one-dimensional field metric is just $d\ell^2=d\varphi^2$. Thus the derivatives below are ordinary derivatives with respect to the canonical field.
\subsection{Further refined de Sitter conjecture}
For the further refined de Sitter condition we use the form \cite{AndriotRoupec2019}
\begin{equation}
\begin{split}
&\left(\Mp\frac{|U'|}{U}\right)^\mathfrak q
-a\Mp^2\frac{U''}{U}\ge b,\\
&a+b=1,
\qquad a,b>0,
\qquad \mathfrak q>2,
\end{split}
\label{eq:FRD}
\end{equation}
We reserve $\mathfrak q$ for the conjectural exponent and keep $q$ for the power of the quintessence tail.
Define
\begin{equation}
F_1\equiv\Mp\frac{|U'|}{U}=\sqrt{2\epsilon_\mathrm{V}},
\qquad
F_2\equiv\Mp^2\frac{U''}{U}=\eta_\mathrm{V}.
\end{equation}
Then
\begin{equation}
F_1^\mathfrak q-aF_2\ge1-a.
\label{eq:FRDdimensionless}
\end{equation}
Using the first-order slow-roll relations, the two quantities can also be written in terms of the observables:
\begin{equation}
F_1=\sqrt{\frac r8},
\qquad
F_2=\frac12\left(n_\mathrm{s}+\frac{3r}{8}-1\right),
\label{eq:Fobs}
\end{equation}
so that, for $F_2<1$,
\begin{equation}
a\ge a_{\min}=\frac{1-F_1^\mathfrak q}{1-F_2}.
\label{eq:amin}
\end{equation}
Here $a$, $b$, and $\mathfrak q$ are parameters of the conjecture itself. They are not CMB observables. At a chosen pivot, the measured quantities simply tell us how the inequality must be arranged at that point.
The asymptotic question is easier. On the plateau, and also on the inverse-power tail,
\begin{equation}
\frac{|U'|}{U}\rightarrow0,
\qquad
\frac{U''}{U}\rightarrow0.
\end{equation}
When the field takes values the two derivative terms slowly vanish. As a result the left side of Eq.~\eqref{eq:FRDdimensionless}. Eventually reaches zero. In contrast the constant term $b$ stays positive and unchanged in this limit. Therefore the FRDSSC condition cannot be met across the field range. However as will be shown below this does not imply that the FRDSSC condition is broken everywhere and there may be regions where the FRDSSC condition remains true.
As an example we select $\mathfrak q=2.2$. Explore the resulting behavior. Then
\begin{equation}
a_{\min}=\frac{1-F_1^\mathfrak q}{1-F_2},
\end{equation}
with
\begin{equation}
a_{\min}\simeq0.98815,\quad0.98554,\quad0.98222
\end{equation}
for $k=2,4,8$, respectively. Equivalently,
\[
b_{\max}\simeq0.01185,\quad0.01446,\quad0.01778.
\]
At the three pivots the inequality can be arranged, but only very near the limiting boundary \(b\rightarrow0\). These values are a local requirement of the conjecture, not a cosmological measurement of $a$ or $b$.
\subsection{Scalar weak gravity conjecture}
For the scalar weak-gravity test we use the potential-level inequality \cite{Palti2017}
\begin{equation}
(U''')^2\geq\frac{(U'')^2}{M_\mathrm{P}^2}.
\label{eq:SWdim}
\end{equation}
With the canonical normalization it is convenient to define
\begin{equation}
d_\mathrm{n}\equiv M_\mathrm{P}^n\frac{U^{(n)}}{U},
\end{equation}
so that
\begin{equation}
\Delta_{\rm SWGC}\equiv d_3^2-d_2^2\ge0.
\label{eq:SW}
\end{equation}
The physical meaning of the SWGC needs some care. In Palti's original construction the scalar is related to a force between states with field-dependent masses. The potential inequality used in the present literature comes only after making an additional identification with a second derivative of the potential. Gonzalo and Ib\'{a}\~nez pointed out the limitation of applying that statement directly to an arbitrary scalar and proposed the strong version for that case. For this reason we use Eq.~\eqref{eq:SWdim} as a potential-level diagnostic. We do not interpret it as a direct microscopic force calculation for the inflaton. At the pivot this distinction is particularly clear because $U''<0$.
There is also a convention issue in the numerical sign. In the main calculation we use the reduced $M_\mathrm{P}$. As a check, we replace the mass scale on the right-hand side of the dimensionful inequalities by the unreduced $M_{\mathrm{Pl}}=\sqrt{8\pi}\,M_\mathrm{P}$, without changing the definition of $d_\mathrm{n}$. We then find
\[
\Delta_{\rm SWGC}
=(5.21\times10^{-6},\,7.69\times10^{-6},\,1.70\times10^{-5}),
\]
and
\[
\Delta_{\rm SSWGC}
=(4.81\times10^{-6},\,1.26\times10^{-5},\,3.27\times10^{-5})
\]
for $k=2,4,8$, respectively. Neither choice is preferred by the calculation itself. We keep the reduced-$M_\mathrm{P}$ convention in the main text and show the second one only because it makes the normalization dependence explicit.

\subsection{Strong scalar weak gravity conjecture}
The SSWGC is given by \cite{GonzaloIbanez2019}
\begin{equation}
2(U''')^2-U''U''''\geq\frac{(U'')^2}{\Mp^2},
\label{eq:SSWdim}
\end{equation}
or
\begin{equation}
\SSW\equiv2d_3^2-d_2d_4-d_2^2\ge0.
\label{eq:SSW}
\end{equation}
The SSWGC involves another combination of the same derivatives; it is not simply the SWGC with a different numerical coefficient \cite{GonzaloIbanez2019}. Because $d_3$, $d_4$, and $d_2$ enter differently, the two tests can give different answers at a finite field value. In the far plateau limit the higher-derivative terms fall faster than $d_2^2$, and the negative asymptotic sign follows.

\section{Analytical limits of the scalar swampland tests}
\label{sec:limits}

\subsection{Inflationary plateau}
On the inflationary side, taking $y=-\varphi\gg M$, define
\begin{equation}
A\equiv\frac{M_\mathrm{P}^2}{8\alpha\lambda^k}.
\end{equation}
Then
\begin{equation}
U=\frac{M_\mathrm{P}^2}{8\alpha}\left(1-Ay^{-k}+\cdots\right).
\end{equation}
The leading canonical derivatives are
\begin{equation}
\begin{split}
d_2&\sim-k(k+1)A M_\mathrm{P}^2 y^{-k-2},\\
d_3&\sim-k(k+1)(k+2)A M_\mathrm{P}^3 y^{-k-3},\\
d_4&\sim-k(k+1)(k+2)(k+3)A M_\mathrm{P}^4 y^{-k-4}.
\end{split}
\end{equation}
The large-field hierarchy is already visible in the ratio
\begin{equation}
\frac{d_3^2}{d_2^2}
\sim\frac{(k+2)^2M_\mathrm{P}^2}{y^2}\longrightarrow0,
\end{equation}
so $\Delta_{\rm SWGC}$ eventually turns negative. At the same time,
\begin{equation}
F_1\sim y^{-k-1}\rightarrow0,
\qquad
F_2\sim y^{-k-2}\rightarrow0.
\end{equation}
We see that the FRDSSC reaches the obstacle because its two field‑derivative inputs vanish.
We do not need any numerical root to support this statement. The power‑law method to the plateau already forces both $U''/U$ and $d_3^2/d_2^2$ to drop when $y$ becomes large.
The negative outcome, at infinity is fixed by the shape of the potential. The small intervals that show up in the numerical scan depend on the way the interpolation is carried out.
\subsection{Inverse-power quintessence tail}
On the quintessence side, for $\varphi\gg M$, we obtain
\begin{equation}
U(\varphi)\simeq C_\mathrm{q}\varphi^{-q},
\qquad
C_\mathrm{q}\equiv\lambda^kM^{k+q}.
\end{equation}
For this branch it is easiest to work with the ratios of derivatives to the potential:
\begin{align}
\frac{U'}U&\simeq-\frac q\varphi,\\
\frac{U''}U&\simeq\frac{q(q+1)}{\varphi^2},\\
\frac{U'''}U&\simeq-\frac{q(q+1)(q+2)}{\varphi^3},\\
\frac{U''''}U&\simeq\frac{q(q+1)(q+2)(q+3)}{\varphi^4}.
\end{align}
This gives\begin{equation}
\SW\simeq
\frac{q^2(q+1)^2\Mp^6}{\varphi^6}
\left[(q+2)^2-\frac{\varphi^2}{\Mp^2}\right],
\label{eq:SWtail}
\end{equation}
and
\begin{equation}
\SSW\simeq
\frac{q^2(q+1)^2\Mp^6}{\varphi^6}
\left[(q+1)(q+2)-\frac{\varphi^2}{\Mp^2}\right].
\label{eq:SSWtail}
\end{equation}
The signs can then be read off directly. The asymptotic zeroes occur at
\[
\varphi_{\rm SWGC}^{(0)}=(q+2)M_\mathrm{P},
\quad
\varphi_{\rm SSWGC}^{(0)}=\sqrt{(q+1)(q+2)}\,M_\mathrm{P}.
\]
For $q=k=2,4,8$ these are
$
\varphi_{\rm SWGC}^{(0)}=(4,6,10)M_\mathrm{P},
$
and
$
\varphi_{\rm SSWGC}^{(0)}=(\sqrt{12},\sqrt{30},\sqrt{90})M_\mathrm{P}.
$
These numbers belong to the asymptotic formulas. They should not be confused with the exact zeroes of the finite-$\alpha$ expressions used in the numerical plots. The same tail also leaves the refined de Sitter obstruction in place. In the limit under consideration,
\begin{equation}
F_1\simeq q\frac{\Mp}{\varphi}\rightarrow0,
\qquad
F_2\simeq q(q+1)\frac{\Mp^2}{\varphi^2}\rightarrow0.
\end{equation}
The asymptotic tail tells us the eventual sign of the diagnostic, but it does not determine where the cosmological field stops rolling. That requires the late-time dynamics. For $k=q=4$, one can still quote the field scale obtained by matching the asymptotic potential to the present dark-energy density:
\begin{equation}
U(\varphi_\mathrm{match})\simeq
\lambda^4M^8\varphi_\mathrm{match}^{-4}
\simeq\rho_{\rm DE,0},
\end{equation}
which gives
\begin{equation}
\frac{\varphi_\mathrm{match}}{M_\mathrm{P}}\sim12.3.
\end{equation}
The value $12.3M_\mathrm{P}$ should be regarded only as a matching scale. It is not the output of a dynamical freezing calculation. To obtain such a result one would have to follow the post-inflationary evolution through reheating and kination and impose the corresponding late-time constraints, including the gravitational-wave and BBN requirements. We do not perform that analysis here \cite{Dimopoulos2025}.
For $q=4$, the asymptotic sign changes are at
\begin{equation}
\varphi=6M_\mathrm{P},
\qquad
\varphi=\sqrt{30}\,M_\mathrm{P},
\end{equation}
For the SWGC and SSWGC. The coincidence scale lies beyond both leading sign changes. The asymptotic diagnostics are already negative there. The comparison is useful. The comparison does not locate the actual freezing solution. The same caution applies to $k=2$ and $k=8$ which remain benchmark trajectories unless $k=2$ and $k=8$ have late-time normalization and evolution imposed separately.

\section{Distance conjecture}
\label{sec:distance}
For the canonical one-dimensional field space there is no nontrivial geodesic calculation to perform. The distance is simply \cite{OoguriVafa2007,Etheredge2022,LeeLercheWeigand2022}
\begin{equation}
\Delta\varphi=|\varphi_2-\varphi_1|.
\end{equation}
The important part of the Distance Conjecture is not the distance formula itself, but the ultraviolet statement that an appropriate infinite-distance direction should bring down an exponentially light tower,
\begin{equation}
m_{\rm tower}(\varphi)
\simeq m_{\rm ref}\exp\left[-c\frac{|\varphi-\varphi_{\rm ref}|}{M_\mathrm{P}}\right],
\quad c=\mathcal O(1),
\end{equation}
provided the scalar direction can indeed be identified with the relevant modulus or scalar direction of the ultraviolet completion.
Our effective action contains neither the tower nor a preferred point $\varphi_{\rm ref}$. Consequently the super-Planckian excursion is a statement about the trajectory, not a direct SDC exclusion. A numerical value of $c$ would also require extra ultraviolet input. The last column of Table~\ref{tab:sdctcc} is therefore only a diagnostic of a hypothetical tower.

\section{Trans-Planckian censorship conjecture}
\label{sec:TCC}
The TCC asks a different question, namely how much accelerated expansion the solution contains \cite{BedroyaBrandenbergerLoverdeVafa2019,BedroyaVafa2020}:
\begin{equation}
N_{\mathrm{TCC}}\equiv\int_{t_\mathrm{i}}^{t_\mathrm{f}}H\,dt
\lesssim\ln\left(\frac{M_\mathrm{P}}{H_\mathrm{f}}\right),
\label{eq:TCC}
\end{equation}
up to convention-dependent order-one factors. For a quasi-de Sitter stage with slowly varying $H$, this becomes
\begin{equation}
N_{\mathrm{TCC}}\lesssim\ln\left(\frac{M_\mathrm{P}}{H}\right).
\label{eq:TCCsimple}
\end{equation}
If the pivot exits after inflation has begun, the complete accelerated interval is no shorter than the pivot-to-end segment, so $N_{\mathrm{TCC}}\ge N_*$. The comparison with the logarithmic upper estimate can then be made directly.
For the three numerical solutions we find
\begin{equation}
\ln\left(\frac{M_\mathrm{P}}{H_{\mathrm{end}}}\right)
\simeq12.43,\quad13.36,\quad15.17,
\end{equation}
We take $N_*=60$ as the main case and $N_*=70$ as a sensitivity check. In either case the difference is large: the logarithmic estimate gives only about $12$-$15$ e-folds, compared with the inflationary interval of interest here.
The frequently quoted energy-scale form, $V^{1/4}\lesssim6\times10^6\,\textrm{GeV}$ in the analysis of Bedroya et al., uses additional assumptions about the post-inflationary history. It is therefore not a universal, reheating-independent bound. This matters for quintessential inflation because kination changes the relation between the end of inflation and the observable pivot. For the present benchmark the direct e-fold comparison is the cleaner diagnostic.
For the quintessence-side scans we set $q=k$; this choice does not enter the inflationary branch. The inflationary
observables on the $\varphi<0$ branch remain independent of $q$.

\section{Numerical confrontation with observations}
\label{sec:numerics}
For the numerical benchmark we solve the exact system with
\begin{equation}
\begin{split}
&A_\mathrm{s}=2.1\times10^{-9},
\qquad
M=1.38\times10^{-13}\Mp,\\
&\alpha=2\times10^8,
\end{split}
\end{equation}
The three unknown quantities $\lambda$, $\varphi_{\rm end}$, and $\varphi_*$ are fixed by the amplitude, the end-of-inflation condition, and the e-fold integral. We retain $M=1.38\times10^{-13}M_\mathrm{P}$, the value adopted in Ref.~\cite{Dimopoulos2025}; it is not adjusted here to fit the late-time dark-energy density. The benchmark is likewise not a complete reheating or BBN analysis. As a check on the normalization, the earlier quadratic Palatini calculation gives approximately $n_\mathrm{s}=0.966$ and $r=0.017$ for the $k=4$, $N_*=60$ case \cite{Dimopoulos2025}.
\begin{table*}[t]
\centering
\caption{Exact benchmark solutions in reduced Planck units for the three inflationary exponents.}
\label{tab:bench}
\small
\begin{ruledtabular}
\begin{tabular}{c
                S[table-format=1.3e-1]
                S[table-format=-1.3]
                S[table-format=-2.3]
                S[table-format=1.3e-1]
                S[table-format=-1.3e-1]
                S[table-format=1.5]
                S[table-format=1.5]
                S[table-format=2.3]}
\multicolumn{1}{c}{$k$} &
\multicolumn{1}{c}{$\lambda$} &
\multicolumn{1}{c}{$\varphi_{\rm end}$} &
\multicolumn{1}{c}{$\varphi_*$} &
\multicolumn{1}{c}{$\epsilon_*$} &
\multicolumn{1}{c}{$\eta_*$} &
\multicolumn{1}{c}{$n_\mathrm{s}$} &
\multicolumn{1}{c}{$r$} &
\multicolumn{1}{c}{$\Delta\varphi_*/\Mp$}\\
\hline
\noalign{\vskip 2pt} 
2 & 5.519e-6 & -1.306 & -9.010 & 1.003e-3 & -1.090e-2 & 0.97218 & 0.01605 & 7.705\\
4 & 5.931e-4 & -2.795 & -13.002 & 1.068e-3 & -1.350e-2 & 0.96659 & 0.01708 & 10.207\\
8 & 4.577e-3 & -5.655 & -19.888 & 1.109e-3 & -1.688e-2 & 0.95959 & 0.01775 & 14.233\\
\end{tabular}
\end{ruledtabular}
\end{table*}
The $k=4$ entry agrees with the previously reported benchmark. When $k$ is raised from $2$ to $8$, the tilt moves toward smaller values, while $r$ changes only mildly. All three tensor predictions stay below the representative limits quoted in Table~\ref{tab:obs}.
\begin{table*}[t]
\centering
\caption{Comparison with observational constraints. The scalar-index entries use the P-ACT-LB value $n_\mathrm{s}$ equals 0.9743 plus or minus 0.0034; tensor limits are listed separately for the BICEP Keck combinations. The quoted significances are one comparisons, with the stated central values and uncertainties and do not represent a full likelihood analysis.}
\label{tab:obs}
\small
\begin{ruledtabular}
\begin{tabular}{l c c c c c}
Dataset & $n_\mathrm{s}$ benchmark & $r$ bound & $k=2$ & $k=4$ & $k=8$\\
\hline
\noalign{\vskip 2pt}
Planck 2018 & $0.9649\pm0.0042$ & $-$ & $1.73\sigma$ & $0.40\sigma$ & $1.27\sigma$\\
P-ACT-LB & $0.9743\pm0.0034$ & $-$ & $0.62\sigma$ & $2.27\sigma$ & $4.33\sigma$\\
P-ACT-LB-BK18 & $-$ & $<0.038$ & within bound & within bound & within bound\\
BICEP/Keck 2018 & $-$ & $<0.036$ & within bound & within bound & within bound\\
\end{tabular}
\end{ruledtabular}
\end{table*}
At $N_*=60$ the scalar tilt gives the clearest separation between the three cases. The $k=2$ result is near the P-ACT-LB central value, the $k=4$ point is closer to the Planck value, and the $k=8$ point is redder than P-ACT-LB. These are still only pivot comparisons, and the chosen value of $N_*$ remains tied to the post-inflationary history.
\subsection{FRDSSC at the pivot}
As one example, we take $\mathfrak q=2.2$. Then
\begin{equation}
a_{\min}=\frac{1-(r/8)^{1.1}}
{1-\tfrac12(n_\mathrm{s}+3r/8-1)}.
\end{equation}
For the benchmark points,
\begin{table}[t]
\centering
\caption{FRDSSC requirement at the CMB pivot for the illustrative choice $\mathfrak q=2.2$.}
\label{tab:frd}
\small
\begin{ruledtabular}
\begin{tabular}{c c c c}
$k$ & $F_1$ & $F_2$ & $a_{\min}$\\
\hline
\hline
\noalign{\vskip 2pt}
2 & 0.04479 & $-1.090\times10^{-2}$ & 0.98815\\
4 & 0.04621 & $-1.350\times10^{-2}$ & 0.98554\\
8 & 0.04710 & $-1.688\times10^{-2}$ & 0.98222\\
\end{tabular}
\end{ruledtabular}
\end{table}
Table~\ref{tab:frd} shows that the condition can be arranged at each pivot. The price is that $a$ has to lie very close to one, or equivalently $b$ has to be small. These numbers are not measurements of the conjectural parameters.
\subsection{SWGC and SSWGC at the pivot}
Evaluating the finite-$\alpha$ derivatives of Eqs.~\eqref{eq:U1}-\eqref{eq:U4} at the pivot gives
\begin{table*}[t]
\centering
\caption{Potential-level SWGC and SSWGC diagnostics at the CMB pivot. Negative entries mean that the corresponding inequality fails in the reduced-$M_\mathrm{P}$ convention used in the main analysis.}
\label{tab:swgc}
\small
\begin{ruledtabular}
\begin{tabular}{c c c c c c}
$k$ & $d_2$ & $d_3$ & $d_4$ & $\SW$ & $\SSW$\\
\hline
\hline
\noalign{\vskip 2pt}
2 & $-1.0900\times10^{-2}$ & $-3.1517\times10^{-3}$ & $-9.4766\times10^{-4}$ &
$-1.0888\times10^{-4}$ & $-1.0927\times10^{-4}$\\
4 & $-1.3500\times10^{-2}$ & $-3.8656\times10^{-3}$ & $-7.4534\times10^{-4}$ &
$-1.6731\times10^{-4}$ & $-1.6243\times10^{-4}$\\
8 & $-1.6878\times10^{-2}$ & $-5.3208\times10^{-3}$ & $-7.4757\times10^{-4}$ &
$-2.5657\times10^{-4}$ & $-2.4088\times10^{-4}$\\
\end{tabular}
\end{ruledtabular}
\end{table*}
In the reduced-$M_\mathrm{P}$ convention the two discriminants are negative at all three pivot points. With the alternative Planck-scale normalization their signs become positive. Thus the pivot failure is a statement within the convention used in the main calculation, not a convention-independent exclusion.
\begin{figure*}[t]
\centering
\includegraphics[width=\linewidth]{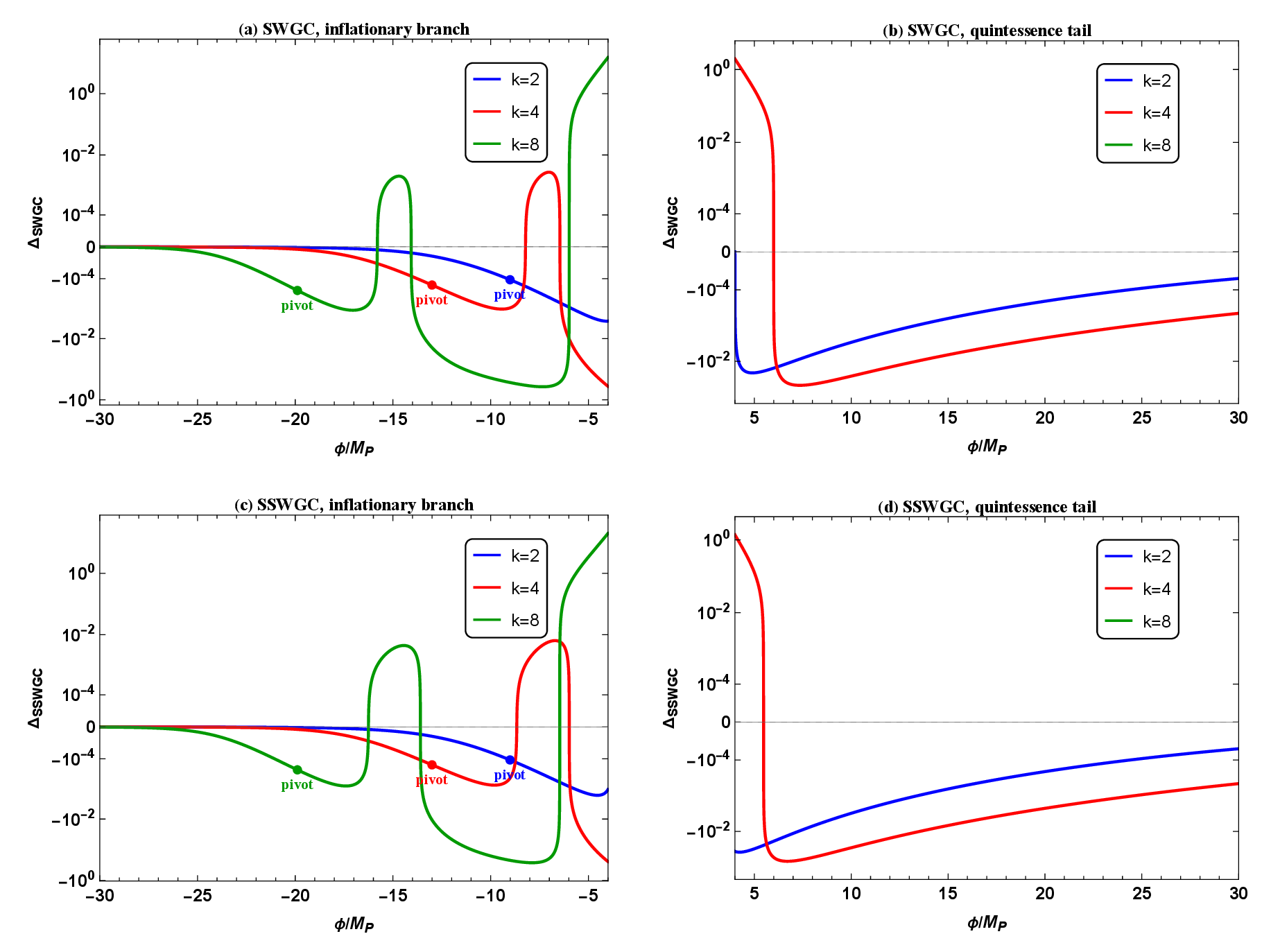}
\caption{Scalar weak-gravity diagnostics as functions of the canonical field $\varphi/M_\mathrm{P}$ for $k=2,4,8$ with $q=k$. The full ranges for both branches are displayed with no interval left out near the transition. For $k=q=2$ and $4$, the value $\varphi=0$ is left out of the SSWGC because $U''''$ is not defined there. The CMB pivots are indicated. The exact SSWGC scan has intervals while the pivot values are negative, in the reduced-$M_\mathrm{P}$ approach used in the main study}
\label{fig:swgc-sswgc}
\end{figure*}
The scan in Fig.~\ref{fig:swgc-sswgc} also shows why a pivot-only statement is insufficient. The full inflationary path contains regions where the sign changes. Both branches are displayed without cutting out a finite neighborhood of the matching point. Only the single point $\varphi=0$ is left out of the SSWGC for $k=q=2$ and $4$, because there $U''''$ does not exist.
The positive parts of the exact scan can be listed explicitly. With $y=-\varphi>0$, the SWGC gives
$
1.47<y<3.28\hspace{1.5pt}(k=2),\hspace{1.5pt}
6.47<y<8.24\hspace{1.5pt}(k=4),\hspace{1.5pt}
14.16<y<15.89\hspace{1.5pt}(k=8).
$
If the same intervals are expressed as e-folds measured from the end of inflation, one finds
$
0.12<N<2.95,\hspace{1.5pt}
4.86<N<10.09,\hspace{1.5pt}
11.78<N<17.73,
$
respectively. The corresponding inflection points, defined by $U''=0$, are
approximately
$
y=2.62,\hspace{1.5pt} 7.42,\hspace{1.5pt} 14.98.
$
The corresponding SSWGC scan gives
$
1.31<y<3.74,\hspace{1.5pt}
5.98<y<8.66,\hspace{1.5pt}
13.61<y<16.24
$
for $k=2,4,8$. Measured in e-folds from the end of inflation, they become
$
0<N<4.24,\hspace{1.5pt}
3.87<N<11.89,\hspace{1.5pt}
10.42<N<19.42.
$
These are field features of the exact expressions. They do not change the sign at the pivot. They disappear again in the large-field limits.
There is one detail for $k=8$. Immediately after the end of inflation the scan shows another narrow positive interval.
$
5.655<y<5.99
$
for the SWGC and
$
5.655<y<6.48
$
for the SSWGC. So, in the $k=8$ case there is more than one positive piece in the scan.
\subsection{Distance and TCC diagnostics}
All three benchmark excursions in Table~\ref{tab:bench} are larger than $M_P$. The associated Hubble scales lead to
\begin{table*}[t]
\centering
\caption{Canonical field excursion and TCC diagnostics for the three benchmark solutions. The last column is an illustrative tower-suppression parameter, not an SDC bound.}
\label{tab:sdctcc}
\small
\begin{ruledtabular}
\begin{tabular}{c c c c c c}
$k$ & $H_*/\Mp$ & $H_{\mathrm{end}}/\Mp$ & $\ln(\Mp/H_{\mathrm{end}})$ & $U_{\rm end}^{1/4}$ [GeV] & $c_{\max}$ from $m_{\rm tower}>H_*$\\
\hline
\hline
\noalign{\vskip 2pt}
2 & $1.290\times10^{-5}$ & $4.00\times10^{-6}$ & 12.43 & $6.41\times10^{15}$ & 1.46\\
4 & $1.331\times10^{-5}$ & $1.58\times10^{-6}$ & 13.36 & $4.02\times10^{15}$ & 1.10\\
8 & $1.356\times10^{-5}$ & $2.59\times10^{-7}$ & 15.17 & $1.63\times10^{15}$ & 0.79\\
\end{tabular}
\end{ruledtabular}
\end{table*}
where
\begin{equation}
c_{\max}\equiv\frac{\ln(\Mp/H_*)}{\Delta\varphi_*/\Mp}.
\end{equation}
The last column should not be read as an SDC bound. It only asks what suppression an assumed tower with an order-one exponential coefficient would acquire over the calculated distance.
For $N_*=60$ or $70$,
\begin{equation}
N_*\gg\ln(\Mp/H_{\mathrm{end}})
\end{equation}
for every benchmark. With the standard quasi-de Sitter interpretation, the required inflationary interval is much longer than this estimate.

\section{Consolidated theoretical-observational assessment}
\label{sec:summarytable}
For convenience, Table~\ref{tab:summary} puts the principal numbers side by side. The word local in the FRDSSC column means that the chosen pivot admits some value of $a$. The SWGC and SSWGC columns use the reduced-$M_P$ convention. The SDC entry gives the field distance rather than a pass/fail decision, since no tower has been specified. The TCC statement uses the quasi-de Sitter estimate discussed above.
\begin{table*}[t]
\centering
\caption{Summary of the inflationary, observational, and scalar-swamp-land diagnostics for the three benchmark choices.}
\label{tab:summary}
\small
\begin{ruledtabular}
\begin{tabular}{c c c c c c c c}
$k$ & $n_\mathrm{s}$ & $r$ & ACT tilt & FRDSSC & SWGC & SSWGC & SDC/TCC\\
\hline
\hline
\noalign{\vskip 2pt}
2 & 0.97218 & 0.01605 & $0.62\sigma$ & local & fail & fail &
$\Delta\varphi=7.71\,\Mp$; exceeds standard TCC estimate\\
4 & 0.96659 & 0.01708 & $2.27\sigma$ & local & fail & fail &
$\Delta\varphi=10.21\,\Mp$; exceeds standard TCC estimate\\
8 & 0.95959 & 0.01775 & $4.33\sigma$ & local & fail & fail &
$\Delta\varphi=14.23\,\Mp$; exceeds standard TCC estimate\\
\end{tabular}
\end{ruledtabular}
\end{table*}
The pattern is fairly clear. The tensor amplitudes satisfy the representative bounds, while the scalar index is more sensitive to $k$. The negative weak-gravity entries belong to the reduced-$M_\mathrm{P}$ convention. The SDC number is a distance, not a verdict, and the TCC comparison is conditional on the quasi-de Sitter assumption.
We take $N_*=60$ for the main benchmark. In quintessential inflation this number is connected with the post-inflationary history, especially reheating and kination, so we also repeat the calculation with $N_*=70$. This gives
$
(n_\mathrm{s},r)=
(0.97588,0.01559),\hspace{1.5pt}
(0.97077,0.01674),\hspace{1.5pt}
(0.96422,0.01750)
$
for $k=2,4,8$, respectively. Relative to
$n_\mathrm{s}=0.9743\pm0.0034$, these correspond to
$
0.46\sigma,\hspace{1.5pt}1.04\sigma,\hspace{1.5pt}2.96\sigma.
$
The $N_*=70$ numbers are included to show the sensitivity to the assumed history. They are not evidence that one branch is preferred. Such a statement would require a likelihood analysis over the full parameter set.
\section{Relation to kination and the quintessential tail}
\label{sec:kination}
The evolution after inflation is different from the usual oscillatory reheating picture. The scalar does not have to settle into a quadratic minimum. When its potential energy becomes small compared with its kinetic energy, a kination stage can appear,
\begin{equation}
\rho_\varphi\simeq\frac12\dot\varphi^2\propto a^{-6},
\end{equation}
corresponding to
\begin{equation}
w_\varphi\simeq1.
\end{equation}
Radiation redshifts more slowly than the kinetic component and eventually catches up, after which the usual hot Big Bang evolution begins.
Kination enters even before one discusses the late-time tail, because it changes the relation between the pivot mode and the end of inflation. This is why different post-inflationary histories can give values of $N_*$ in the $60$-$70$ range \cite{Dimopoulos2025}.
Nothing in the scalar diagnostics changes the kination mechanism itself. They do show, however, that the inverse-power tail does not cure the large-distance behavior of the derivative criteria. Since the same canonical field is used on both sides of the transition, the long excursion also makes the Distance Conjecture relevant. To turn that into an actual UV test one would need to know the tower of states and its coupling to the scalar, which are not part of the present effective action.
\section{Interpretation and caveats}
\label{sec:discussion}
There are three separate pieces in the analysis. The first is the Einstein-frame theory obtained from the Palatini action. The second is the inflationary benchmark extracted from that theory. The third consists of the Swampland diagnostics. The first part is exact within the model, whereas the last part is necessarily more conditional.
The benchmark is also not parameter free. The parameter $\alpha$ does not change $V_J$ itself; instead it changes how $V_J$ appears in the Einstein frame,
\begin{equation} \begin{split}
V_\mathrm{J}\longrightarrow\frac{V_\mathrm{J}}{1+8\alpha V_\mathrm{J}/M_\mathrm{P}^2},
\end{split} \end{equation}
and fixes the asymptotic plateau height
\begin{equation} \begin{split}
U_{\rm pl}=\frac{M_\mathrm{P}^2}{8\alpha}.
\end{split} \end{equation}
The exponent $k$ mainly fixes the inflationary shape, while $M$, $\lambda$, $q$, and the chosen value of $N_*$ affect the normalization and the field positions. The table should therefore be viewed as a benchmark slice of parameter space rather than as a parameter-free prediction for the whole quintessential history.
The FRDSSC is already conditional at the pivot because the benchmark requires small $b$. At large field distance a fixed positive $b$ is incompatible with the vanishing derivative ratios. The SWGC and SSWGC give another view of the potential, but their numerical signs depend on the Planck-scale convention used in the dimensionful inequalities.
The two weak-gravity criteria should not be treated as identical. Palti's construction is tied to scalar forces mediated between states with field-dependent masses, while the SSWGC was introduced as a condition on a general scalar sector. For the present purpose the SSWGC is consequently the cleaner potential-level test; the original SWGC carries the extra interpretive step described above.
The excursion is unambiguously super-Planckian. What remains open is what that distance means for the SDC. A large field range becomes a genuine test only when the corresponding ultraviolet tower is known. No such tower is specified in the present action.
The TCC gives another conditional result. The benchmark has $H_*\sim10^{-5}M_\mathrm{P}$, but the corresponding logarithmic estimates at the end of inflation are only about $10$-$15$ e-folds. The required $60$-$70$ e-fold interval is much longer. This is a tension with the standard TCC translation used here, not a proof that the effective theory has no ultraviolet completion.

\section{Conclusions}
\label{sec:conclusions}
We have studied the quadratic Palatini $F(R,X)$ model from its Einstein-frame formulation, through the inflationary phase, and finally along the same scalar trajectory toward the scalar swampland tests. After introducing the auxiliary field, the gravitational sector reduces to Einstein gravity coupled to a canonical scalar, with the effective potential
\begin{equation}
U(\varphi)=
\frac{V_\mathrm{J}(\varphi)}
{1+8\alpha V_\mathrm{J}(\varphi)/M_\mathrm{P}^2}.
\end{equation}
The negative-field side approaches a finite plateau, whereas the positive-field side keeps the inverse-power behaviour inherited from the generalized Peebles-Vilenkin potential. Thus, the inflationary and quintessence regimes can be discussed in terms of one canonical field rather than two separate sectors. For $A_\mathrm{s}=2.1\times10^{-9}$, $M=1.38\times10^{-13}M_\mathrm{P}$, $\alpha=2\times10^8M_\mathrm{P}^{-2}$, and $N_*=60$, we obtain
$n_\mathrm{s}=0.97218,\hspace{1.5pt}0.96659,\hspace{1.5pt}0.95959$
and
$r=0.01605,\hspace{1.5pt}0.01708,\hspace{1.5pt}0.01775$
for $k=2,4,8$, respectively. The $k=4$ case agrees with the benchmark quoted previously. 

In all three cases the tensor contribution remains below the representative bounds adopted here. The scalar tilt, however, shows a much clearer dependence on k. It also changes noticeably when $N_*$ is increased from $60$ to $70$. This makes the assumed post-inflationary evolution relevant to the observational comparison, rather than a detail that can simply be ignored.

Our results show how the Palatini formulation affects the model considered here. The quadratic curvature correction generates a plateau in the inflationary potential, and the predicted tensor-to-scalar ratios remain below the representative observational limits used in our analysis. At \(N_*=60\), the \(k=2\) case gives the closest value of the scalar spectral index to the P-ACT-LB central value. The agreement is less favourable for \(k=4\) and \(k=8\), which deviate more from that value. These findings indicate that the Palatini setup can provide a viable description of the observations for suitable choices of the model parameters. They do not, however, imply that the metric formulation is incompatible with ACT data. To establish whether the two formulations differ significantly in their observational predictions, the metric counterpart must be worked out and tested against the same data, using consistent assumptions about the post-inflationary evolution.

The swampland analysis does not lead to a single pass-or-fail conclusion. For $\mathfrak q=2.2$, the further-refined de Sitter condition can be fulfilled at the inflationary pivots, but only when a is close to unity, corresponding to a small b. Once the field is taken sufficiently far from the transition, both $|U'|/U$ and $U''/U$ approach zero. The same condition therefore ceases to hold for any fixed $b>0$.
At the inflationary pivots, the reduced-$M_\mathrm{P}$ SWGC and SSWGC discriminants are both negative. This sign, however, is not maintained over the entire trajectory. The exact numerical scans show finite regions in which the discriminants become positive, followed again by negative values at sufficiently large field distance. The pivot result should consequently not be identified with the behaviour of the full solution. The conclusion also depends on the precise SWGC convention and on how these conjectures are interpreted.

The canonical field excursion is super-Planckian for all three benchmark choices. We do not interpret this by itself as a violation of the distance conjecture, since the corresponding ultraviolet tower of states has not been specified. For $k=q=4$, the asymptotic quintessence tail can nevertheless be matched to the present dark-energy density, which gives
$\varphi_\mathrm{F}\sim12.3M_\mathrm{P}$. This value lies beyond the leading sign-change scales of the two derivative-based diagnostics. It should not, however, be regarded as evidence for a dynamically frozen solution. Establishing such a late-time state requires the post-inflationary evolution, in particular reheating and kination, to be followed explicitly.

The trans-Planckian censorship conjecture provides a separate conditional test. For the three numerical solutions we find
$\ln(M_\mathrm{P}/H_{\mathrm{end}})\simeq12.43,\hspace{1.5pt}13.36,\hspace{1.5pt}15.17$.
These values are well below the standard logarithmic estimate corresponding to a $60-70$ e-fold inflationary interval. Within the quasi-de Sitter interpretation adopted here, the benchmark therefore shows a clear tension with the TCC estimate. This should not be read as proof that no ultraviolet completion exists. The statement is more limited: under the assumptions entering the present benchmark, the corresponding TCC condition is not satisfied.
Taken together, the results are best viewed as several separate pieces of evidence rather than as one overall verdict. The inflationary solution possesses the required plateau and gives tensor amplitudes below the representative observational limits used in this work. The scalar derivative criteria reveal both local and asymptotic tensions, but their interpretation depends on the precise conjecture and on the convention chosen for the Planck mass. The distance conjecture cannot be assessed decisively without specifying an ultraviolet tower, while the TCC conclusion depends on the treatment of the post-inflationary history. 

As the future works, the most direct next step is to determine the post-inflationary history instead of treating $N_*$ as an external input. Perturbative and non-perturbative reheating, followed by the kination era, could be evolved explicitly and would fix the pivot location within the same calculation that describes the transition to quintessence. A full Bayesian analysis of ACT DR6, SPT, Planck, and BICEP/Keck data would provide a more quantitative way of separating the $k=2,4,8$ cases. On the ultraviolet side, a string-motivated construction containing an explicit tower of states would make the distance-conjecture test considerably more concrete. Introducing a gauge sector with charged states would likewise allow an independent WGC analysis. Other natural directions include more general Palatini curvature functions, noncanonical kinetic structures, multifield extensions, reheating observables, primordial gravitational waves, and magnetogenesis
\begin{acknowledgments}
The work of Kazuharu Bamba was supported in part by the JSPS KAKENHI Grant
No.~25KF0176 and a grant-in-aid of academic research of Yamaguchi Scholarship
Foundation.
\end{acknowledgments}
\appendix
\section{Detailed derivatives on the inflationary branch}
\label{app:inflationderivatives}
In this appendix, we collect the derivatives needed for the inflationary-branch calculations. The inflationary branch derivatives are gathered here for reference. These derivatives serve as the algebraic input, for the exact Einstein-frame derivatives. They are also used in gravity and asymptotic calculations.
For $\varphi\leq0$,
\begin{equation}
V=\lambda^k(\varphi^k+M^k),
\end{equation}
and
\begin{align}
V'&=k\lambda^k\varphi^{k-1},\\
V''&=k(k-1)\lambda^k\varphi^{k-2},\\
V'''&=k(k-1)(k-2)\lambda^k\varphi^{k-3},\\
V''''&=k(k-1)(k-2)(k-3)\lambda^k\varphi^{k-4}.
\end{align}
Using these derivatives in Eqs.~\eqref{eq:U1}-\eqref{eq:U4} gives
\begin{equation}
\begin{split}
\frac{U'}U&=\frac{V'}{VD},\\
\frac{U''}U&=\frac{V''}{VD}-\frac{2C(V')^2}{VD^2},\\
\frac{U'''}U&=\frac{V'''}{VD}-\frac{6CV'V''}{VD^2}
+\frac{6C^2(V')^3}{VD^3},\\
\frac{U''''}U&=\frac{V''''}{VD}-\frac{8CV'V'''}{VD^2}
-\frac{6C(V'')^2}{VD^2}\\
&\quad+\frac{36C^2(V')^2V''}{VD^3}
-\frac{24C^3(V')^4}{VD^4}.
\end{split}
\end{equation}
\section{Derivative relations on the quintessence branch}
\label{app:tail}
In this appendix, we give the derivative relations used on the quintessence branch. For the positive-field branch the first two derivatives have a compact form; the higher derivatives follow by differentiating the same rational function. For
\begin{equation}
V=C_\mathrm{q}(\varphi^q+M^q)^{-1},
\qquad
C_\mathrm{q}=\lambda^kM^{k+q},
\end{equation}
we obtain
\begin{align}
V'&=-C_\mathrm{q} q\varphi^{q-1}(\varphi^q+M^q)^{-2},\\
V''&=C_\mathrm{q} q\varphi^{q-2}
\frac{(q+1)\varphi^q-(q-1)M^q}
{(\varphi^q+M^q)^3}.
\end{align}
The third derivative and the fourth derivative are obtained by carrying out one differentiation. With Eqs. \eqref{eq:U1}-\eqref{eq:U4} these expressions give the exact finite‑$\alpha$ input that is used in the main text and reduce to the simple inverse‑power relations when the asymptotic limit is reached.
\section{Slow-roll benchmark calculation}
\label{app:benchmark}
In this appendix, we describe the procedure used to obtain the slow-roll benchmark solutions. For each even $k$, the end of inflation is obtained from $\epsilon_V(\varphi_{\rm end})=1$. The chosen e-fold number then fixes the pivot field, while the scalar amplitude fixes the remaining normalization. We solve
\begin{equation}
\begin{split}
&\epsilon_V(\varphi_{\rm end})=1,\\
&N_* =\frac1{\Mp^2}\int_{\varphi_{\rm end}}^{\varphi_*}
\frac{U}{U'}d\varphi,\\
&A_\mathrm{s}=\frac{U_*}{24\pi^2\Mp^4\epsilon_{V,*}}.
\end{split}
\end{equation}
For $N_*=60$, $A_\mathrm{s}=2.1\times10^{-9}$, $M=1.38\times10^{-13}\Mp$, and $\alpha=2\times10^8$, the numerical roots give Table~\ref{tab:bench}. If we repeat the same procedure for $N_*=70$, we obtain
\[
n_\mathrm{s}=0.97588,\quad0.97077,\quad0.96422
\]
and
\[
r=0.01559,\quad0.01674,\quad0.01750
\]
for $k=2,4,8$, respectively. The change is a direct measure of how much the observational comparison depends on the assumed post-inflationary e-fold interval.
\section{FRDSSC data translation}
\label{app:FRD}
In this appendix, we show how the FRDSSC condition is translated into a constraint on the observational parameters. For the observational translation we start from\begin{equation}
F_1^\mathfrak q-aF_2\ge1-a,
\end{equation}
we obtain
\begin{equation}
a(1-F_2)\ge1-F_1^\mathfrak q.
\end{equation}
For $F_2<1$,
\begin{equation}
a\ge\frac{1-F_1^\mathfrak q}{1-F_2}.
\end{equation}
At the illustrative ACT-era model point $n_\mathrm{s}=0.9743$ and $r=0.017$,
\begin{equation}
F_1\simeq0.04610,
\qquad
F_2\simeq-0.00966,
\end{equation}
and for $\mathfrak q=2.2$,
\begin{equation}
a_{\min}\simeq0.9893,
\qquad
b_{\max}\simeq0.0107.
\end{equation}
\vspace{10pt}Again, this is a condition imposed at one selected point of the model. It is not a measurement of $a$ or $b$.

\section{Asymptotic no-go statements}
\label{app:nogo}
In this appendix, we summarize the asymptotic limits relevant to the Swampland diagnostics. At large field distance on either side,
\begin{equation}
\lim_{|\varphi|\rightarrow\infty}F_1=0,
\qquad
\lim_{|\varphi|\rightarrow\infty}F_2=0,
\end{equation}
and
\begin{equation}
\lim_{|\varphi|\rightarrow\infty}\frac{d_3^2}{d_2^2}=0.
\end{equation}
For the SSWGC,
\begin{equation}
\lim_{|\varphi|\rightarrow\infty}
\frac{2d_3^2-d_2d_4}{d_2^2}=0.
\end{equation}
In both limits the higher-derivative contribution becomes subleading and the remaining term is $-d_2^2$. The two weak-gravity diagnostics therefore turn negative. Since $F_1$ and $F_2$ also vanish, a fixed positive $b$ prevents the FRDSSC from holding globally.\\
\section{TCC numerical diagnostic}
\label{app:TCC}
In this appendix, we evaluate the TCC bound for the benchmark inflationary solutions. For the TCC estimate we use
\begin{equation}
H^2\simeq\frac{U}{3\Mp^2},
\end{equation}
and evaluate the Hubble scale both at the pivot and at the end of inflation. For the three benchmark solutions,
\begin{align}
H_*&\simeq(1.29,1.33,1.36)\times10^{-5}\Mp,\\
H_{\mathrm{end}}&\simeq(4.00,1.58,0.259)\times10^{-6}\Mp
\end{align}
for $k=(2,4,8)$. This gives\begin{equation}
N_{\mathrm{max}}^{\mathrm{TCC}}\simeq\ln(\Mp/H_{\mathrm{end}})
\simeq(12.43,13.36,15.17),
\end{equation}
which is well below the $N_*=60$-$70$ interval used for the benchmark.

\end{document}